\documentclass[12pt,subeqn]{article}
\usepackage{amsmath}%
\usepackage{amsfonts}%
\usepackage{amssymb}%
\usepackage{amsthm}
\usepackage{graphicx}
\usepackage{graphics}
\usepackage{color}
\usepackage{subfigure}
\usepackage[a4paper,vmargin={18mm,20mm},hmargin={20mm,20mm}]{geometry}

\DeclareGraphicsExtensions{.bmp}
\DeclareGraphicsExtensions{.jpeg}
\DeclareGraphicsExtensions{.gif}
\DeclareGraphicsExtensions{.wmf}
\DeclareGraphicsExtensions{.pdf}
\DeclareGraphicsExtensions{.eps}
\newtheorem{thm}{Theorem}[section]

\newtheorem{exa}[thm]{Example}
\numberwithin{equation}{section}
\numberwithin{figure}{section}
\numberwithin{table}{section}
\usepackage{setspace}
\usepackage[colorlinks]{hyperref}
\usepackage{color}
\usepackage{braket}
\definecolor{red}{rgb}{0.5,0,0}
\definecolor{green}{rgb}{0,0.5,0}
\definecolor{blue}{rgb}{0,0,0.5}
\hypersetup{ colorlinks,
linkcolor=black,
filecolor=darkgreen,
urlcolor=blue,
citecolor=red }
\makeatletter

\newcommand{\Rmnum}[1]{\expandafter\@slowromancap\romannumeral #1@}
\makeatother

\begin{document}

\title{Spectral zeta functions of a 1D Schr\"odinger problem\\}
\author{Joe Watkins\thanks{E-mail: J.Watkins@kent.ac.uk}\\ \\
\textit{School of Mathematics, Statistics and Actuarial Science,}\\
\textit{University of Kent, Canterbury, CT2 7NF, United Kingdom}}
\date{}
\maketitle

\begin{abstract}
\noindent
\looseness+1 We study the spectral zeta functions associated
to the radial Schr\"odinger problem with potential $V(x)=x^{2M}+\alpha x^{M-1}
+(\lambda^2-1/4)/x^2$. Using the quantum Wronskian equation, we provide
results such as closed-form evaluations for some of the
second zeta functions i.e. the sum over the inverse eigenvalues squared.
Also we discuss how our results can be used to derive relationships
and identities involving special functions, using a particular
${}_5F_4$ hypergeometric series as an example. Our work is then extended
to a class of related ${\cal PT}$-symmetric eigenvalue problems. Using
the fused quantum Wronskian we give a simple method for calculating
the related spectral zeta functions. This method has a number of applications
including the use of the
ODE/IM correspondence to compute the (vacuum) nonlocal integrals of motion
$G_n$ which appear in an associated integrable quantum field theory.
\end{abstract}
\newpage

\newpage

\section{Introduction}

\noindent A common set of models to be studied in quantum mechanics are the anharmonic oscillators defined by the Schr\"odinger equation
\begin{equation} \label{ode0}
-\psi'' +x^{2M}\psi=E\psi.
\end{equation}
\noindent Any eigenfunction must obey the requirement that $\psi\rightarrow 0$ as $x\rightarrow +\infty$ and additionally satisfy either the Dirichlet or Neumann conditions at the origin. The anharmonic oscillators
 include the two exactly-solvable cases of the harmonic oscillator ($M=1$) and infinite square well ($M=\infty$).
 For these two models the eigenvalues can be found in closed
 form, whereas for general $M$ numerical
 methods must be used. The Dirichlet eigenvalues $\{E_k^-\}$ and
 Neumann eigenvalues $\{E_k^+\}$ are associated to the spectral
 zeta function

\begin{equation} \label{genzeta}
Z_\mp(s,M)\equiv \sum_{k=0}^\infty\frac{1}{(E_k^\mp)^s}.
\end{equation}

\noindent WKB estimates show that $E_k^\mp \propto k^\frac{2M}{M+1}$ as $k\rightarrow\infty$ and hence
$Z_\mp(s)$ converges for $\Re(s)>(M+1)/2M$. In the two exactly-solvable cases $Z_\mp(s)$ can be
given in terms of the Riemann zeta function.  However $Z_\mp(s)$ can also be calculated
 for $s\in\mathbb{N}$, the most concise example being given by\footnote{Calculations for
two related zeta functions were originally given in \cite{Vorosquartic} and in \cite{Crandall}.}
\begin{equation} \label{vorosz1}
Z_\mp(1)=\frac{\sigma^{2-2\sigma}\Gamma(\sigma(1\pm \frac{1}{2}))\Gamma(\sigma)\Gamma(\frac{1}{2}-\sigma)}{\sqrt{\pi}\Gamma(1-\sigma(1\mp\frac{1}{2}))},
\end{equation}

\noindent where $M>1$ and

\begin{equation} \label{mudef}
\sigma\equiv \frac{1}{M+1}.
\end{equation}

\noindent The zeta functions $Z_\mp(2)$ can be also be calculated \cite{Vorosquartic},
although these functions involve a ${}_5F_4$ hypergeometric series and no example of a closed form
expression has been given.

\par There has been much study related to these  particular spectral zeta functions \cite{Vorosquartic,Vorosspec0,Vorosspectral,Crandall,Vorosairy},
with one significant result being the determination of `sum rules' which relate together different $Z_\mp(s)$
when $s\in\mathbb{N}$ \cite{Vorosquartic}. An elegant example of this
is found for the quartic anharmonic oscillator, the first few sum rules being
\begin{subequations}\label{quarticsrs}
\begin{align}
\nonumber Z_+(1)=&2Z_-(1) \\
\nonumber 2Z_+(2)=&Z_-(2)+3Z_-(1)^2\\
\nonumber 2Z_+(3)=&9Z_-(1)^3 -Z_-(1)^2-Z_-(3).
\end{align}
\end{subequations}
\noindent The techniques for obtaining the sum rules,
which will be covered later, are the consequence of a wider study known
as exact quantisation, a review of many such problems being found
in \cite{Vorosex}. For a review of the physical applications of spectral
zeta functions in general, we direct the reader toward \cite{Tenszf}.

\par We will consider a generalisation of (\ref{ode0}) given by the
Schr\"odinger equation
\begin{equation} \label{ode}
-\psi'' +\left(x^{2M}+\alpha x^{M-1}+
\frac{\lambda^2-\frac{1}{4}}{x^2}\right)\psi=E\psi
\end{equation}
\noindent with boundary conditions on the half-line.
The eigenfunction criteria is that $\psi\rightarrow 0$ as $x\rightarrow +\infty$,
 with the wavefunction having either of the behaviours
$\psi_-\sim x^{\frac{1}{2}+\lambda}$ or $\psi_+\sim x^{\frac{1}{2}-\lambda}$ as $x\rightarrow 0$.
Such eigenvalue problems have a long and rich history and have been investigated
in many contexts such as quasi-exact solvability \cite{QES1,QES2}, integrable
models \cite{On,Alpha1,Alpha2} and ${\cal PT}$-symmetry
\cite{Bender,Wronskian,Proof,Draft}.

\par The two boundary behaviours $\psi_-$ and $\psi_+$ respectively define two sets of spectra
$\{E_k^-\}$ and $\{E_k^+\}$, which we say are the eigenvalues of the `regular' and `irregular' radial problem\cite{Newton}. To these spectra we associate the zeta functions $Z_\mp(s)$, which for $\lambda=1/2$ are the functions defined in (\ref{genzeta}). However for general $\lambda$ the regular and irregular problem are related by the analytic continuation $\lambda\rightarrow - \lambda$ \cite{On}, giving the zeta-function identity
\begin{equation} \label{ancon}
Z_+(s,\lambda)=Z_-(s,-\lambda).
\end{equation}

\noindent Throughout $\lambda$ will be restricted from taking the values
\begin{equation}  \label{invalidlambda}
\lambda=\pm\frac{1}{2}\left((2m_1+1)(M+1)+\alpha\right) \hspace{4mm} \textrm{and}
\hspace{4mm}  \lambda=\pm\left(m_2+\frac{m_3}{2}(M+1)\right)
\end{equation}
\noindent where $m_1,m_2,m_3\in\mathbb{Z}^+$, although when $\alpha=0$
the condition becomes $2m_3\in\mathbb{Z}^+$ \cite{On}.
The first restriction excludes the possibility that any $E_k^\mp$ could
be equal to zero and the second restriction is to ensure linear independence of
the $\psi_-$ and $\psi_+$ \cite{On}, a condition
which is necessary for later work. Finally, as the eigenvalues are known
to have large-$k$ behaviour $E_k^\mp\propto k^\frac{2M}{M+1}$,
the associated spectral zeta functions $Z_\mp(s)$ will converge
when $\Re (s)>(M+1)/2M$. As we will be studying $Z_\mp(s)$ for $s\geq 1$, we also restrict $M>1$
throughout.

\par This paper will be divided into two sections and will investigate
the radial eigenvalue problems (\ref{ode}).
In Section \ref{seccomp} we
use the well-known method involving Green's functions to calculate $Z_\mp(s)$ when $s\in\mathbb{N}$.
Then we will use the quantum Wronskian equation to derive appropriate sum rules and give examples where $Z_+(2)$ can be written in closed form,
including examples involving the anharmonic oscillators.
 Similar techniques are then used to derive functional relations for two different hypergeometric series which appear in the expressions for $Z_\mp(1)$ and $Z_\mp(2)$. These functional relations can recover many specific, known properties of hypergeometric series, as well as providing new identities.
In Section \ref{secpt} we will extend our work to a class of eigenvalue problems commonly investigated in ${\cal PT}$-symmetric quantum mechanics.
The zeta functions ${\cal Z}_K(s)$ of these problems are studied for $s\in\mathbb{N}$ and found to be constructible
by sum rules involving $Z_\mp(s)$. These expressions for ${\cal Z}_K(s)$ are then used in conjunction with the ODE / IM correspondence \cite{BLZSpec,On,Review} to calculate the vacuum nonlocal integrals of motion $G_n$ which appear in a related quantum integrable field theory \cite{BLZ1,BLZ2}.

\section{Radial Schr\"odinger problems} \label{seccomp}

Given that the eigenvalues $\{E_k^\mp\}$ have no known closed form for general $M$,
naively we might expect that associated spectral zeta functions $Z_\mp(s)$ cannot be given explicitly, except in the exactly-solvable cases.
However when $s\in\mathbb{N}$ the zeta functions
are computed using the process given in \cite{Vorosquartic}.
Although $Z_\mp(s)$ can be defined by analytic continuation
when $s\in\mathbb{Z}^-$ \cite{Vorosspec0}, generally there is no known method to calculate their explicit forms
when $s$ takes general values. Therefore we elect to consider $Z_\mp(s)$ only when $s\in\mathbb{N}$, which we notate $Z_\mp(n)$.
\par For a
Hermitian eigenvalue problem the eigenfunctions are complete and
the Green's function is written as
\begin{equation} \label{greensexpand}
R(E;x,x')=\sum_{k=0}^\infty\frac{\psi_k(x)\psi^*_k(x')}{E_k^--E},
\end{equation}
\noindent where $\{E_k^-\}$ are the associated eigenvalues. As the irregular problem is generally non-Hermitian, the eigenfunctions
are not guaranteed to be complete and consequently the Green's function
cannot be written as in (\ref{greensexpand}). Instead $Z_+(n)$ is defined by analytic continuation as in (\ref{ancon}), which is valid except for $\lambda$ as in (\ref{invalidlambda}).
\par As the eigenfunctions in the Hermitian case are necessarily orthonormal, the zeta functions are calculated by
\begin{equation} \nonumber
Z_-(n)=\int_{\mathbb{R}^+}R(0;x_1,x_2)R(0;x_2,x_3)\dots R(0;x_n,x_1) dx,
\end{equation}
\noindent where $\mathbb{R}^+$ denotes the integration over all
positive space in $n$ dimensions. To evaluate this repeated integral,
$R(0;x,x')$ is decomposed as a combination of $\psi_L$ and $\psi_R$, two
linearly-independent wavefunctions which solve the Schr\"odinger equation
when $E=0$. For a radial problem the two wavefunctions have
specific asymptotic behaviours; $\psi_L$ must obey the boundary conditions
at the origin and we require that $\psi_R\rightarrow 0$ as $x\rightarrow +\infty$.
The Green's function is written as

\begin{equation} \nonumber
R(0;x,x')=\frac{1}{{\cal W}}\psi_L(x_<)\psi_R(x_>)
\end{equation}
\noindent where $x_<\equiv \min(x,x')$, $x_>\equiv \max(x,x')$ and
${\cal W}\equiv \psi_R\psi_L'-\psi_R'\psi_L$, the Wronskian of the two solutions. Thus
\begin{equation} \label{gennested}
Z_-(n)=\frac{n!}{{\cal W}^n}\int_{0<x_1<x_2<\dots<x_n<\infty}\psi_L(x_1)
\psi_R(x_n)\prod_{i=1}^{n-1}
\psi_L(x_i)\psi_R(x_{i+1}) dx_1dx_2\dots dx_n.
\end{equation}

\noindent Now we specialise to the regular radial problem (\ref{ode}),
which is solved for $E=0$ by
\begin{equation} \nonumber
\psi_L = x^\frac{\sigma-1}{2\sigma}M_{-\frac{\sigma\alpha}{2},\sigma\lambda}
(2\sigma x^\frac{1}{\sigma}) \hspace{4mm}
\textrm{and} \hspace{4mm}
\psi_R = x^\frac{\sigma-1}{2\sigma}W_{-\frac{\sigma\alpha}{2},-\sigma\lambda}
(2\sigma x^\frac{1}{\sigma}).
\end{equation}
\noindent Here $M$ and $W$ are the Whittaker functions \cite{Grad} and the Wronskian
of the two solutions is given by
\begin{equation} \nonumber
{\cal W}[\psi_R,\psi_L]=\frac{2\Gamma(1+2\sigma\lambda)}
{\Gamma(\frac{1}{2}+\frac{\sigma\alpha}{2}+\sigma\lambda)}.
\end{equation}
\noindent After making the change of variables $p=2\sigma x_1^\frac{1}{\sigma}$ and
$q=2\sigma x_2^\frac{1}{\sigma}$, we find that
\begin{subequations}\label{ints}
\begin{equation}  \label{int1}
Z_-(1)=\frac{\sigma^{2-2\sigma}\Gamma(\frac{1}{2}+
\frac{\sigma\alpha}{2}+
\sigma\lambda)}{4^\sigma\Gamma(1+2\sigma\lambda)}\int^\infty_0 p^{2\sigma-2}
W_{-\frac{\sigma\alpha}{2},-\lambda\sigma}(p)
M_{-\frac{\sigma\alpha}{2},\lambda\sigma}(p)dp
\end{equation}
and
\begin{equation}
\label{int2} Z_-(2)=\frac{2\sigma^{4-4\sigma}\Gamma^2(\frac{1}{2}+
\frac{\sigma\alpha}{2}+
\sigma\lambda)}{16^\sigma\Gamma^2(1+2\sigma\lambda)}\int^\infty_0 q^{2\sigma-2}
W_{-\frac{\sigma\alpha}{2},-\lambda\sigma}^2(q)
\int^q_0 p^{2\sigma-2}M_{-\frac{\sigma\alpha}{2},\lambda\sigma}^2(p)dpdq.
\end{equation}
\end{subequations}

\noindent The evaluation of these integrals requires known
identities from (7.625) in \cite{Grad}.  The integral (\ref{int1})
can be calculated immediately and found to converge
for $\sigma >1/2$, reproducing the restriction $M>1$ which was
imposed earlier. Completing the integration in (\ref{int2}) is more complicated
and can be handled in a similar way to the
calculation for the anharmonic operators as given in \cite{Vorosquartic}.
Completing the integrations gives the first two zeta functions
\begin{subequations} \label{alphazp}
\begin{equation}  \label{alphazp1}
Z_-(1)=
\frac{\sigma^{2-2\sigma}\Gamma(\frac{1}{2}+\frac{\sigma\alpha}{2}+
\sigma\lambda)\Gamma(2\sigma(1+\lambda))\Gamma(2\sigma)}
{4^\sigma\Gamma(1+2\sigma\lambda)\Gamma(\frac{1}{2}+\frac{\sigma\alpha}{2}
+\sigma(2+\lambda))}
{}_3F_2
{\frac{1}{2}+\frac{\sigma\alpha}{2}+\sigma\lambda,2\sigma(1+\lambda),
2\sigma \choose 1+2\sigma\lambda,\frac{1}{2}+\frac{\sigma\alpha}{2}+\sigma(2+\lambda)}
\end{equation}
\noindent and
\begin{align}
 \nonumber
 Z_-(2)=
\frac{\sigma^{4-4\sigma}}{16^\sigma}
\sum_{l,m,n=0}^\infty
\frac{\Gamma(\frac{1}{2}+\frac{\sigma\alpha}{2}+\lambda\sigma+m)
\Gamma(\frac{1}{2}+\frac{\sigma\alpha}{2}+\lambda\sigma+n)}
{\Gamma(1+2\sigma\lambda+m)
\Gamma(1+2\sigma\lambda+n)m!n!}\frac{1}{(m+n+l+2\sigma(1+\lambda))_{l+1}} \\
\label{alphazp2} \times
G^{22}_{33}
{\frac{1}{2}-\sigma\lambda,\hspace{2mm}\frac{1}{2} +\sigma\lambda,
\hspace{2mm} l+m+n+\frac{\sigma\alpha}{2}+ 2\sigma(2+\lambda) \choose
l+m+n+\sigma(4+\lambda)-\frac{1}{2},l+m+n+\sigma(4+3\lambda)-\frac{1}{2},
-\frac{\sigma\alpha}{2}},
\end{align}
\end{subequations}
where $(a)_b$ is the Pochhammer symbol, ${}_pF_q$ is the generalised
hypergeometric series and $G$ is the Meijer-G function \cite{Grad}. The latter two
functions are both evaluated at $x=1$.
\par By setting $\alpha=0$, the Whittaker functions in (\ref{int1}) and
(\ref{int2}) are replaced by Bessel functions and the integration techniques directly follow those in \cite{Vorosquartic}. The zeta functions
then take on the neater forms
\begin{subequations} \label{azp} \begin{equation} \label{azp1}
Z_-(1,\alpha=0)=
\frac{\sigma^{2-2\sigma}\Gamma(\sigma(1+\lambda))\Gamma(\sigma)
\Gamma(\frac{1}{2}-\sigma)}
{4\sqrt{\pi}\Gamma(1-\sigma(1-\lambda))}
\end{equation}
\noindent and
\begin{align} \label{azp2} \nonumber  Z_-(2,\alpha=0)=
\frac{\sqrt{\pi}\sigma^{3-4\sigma}}
{4^{1+\sigma\lambda}(1+\lambda)} &
\frac{\Gamma(2\sigma(1+\lambda))\Gamma(\sigma(2+\lambda))\Gamma(2\sigma)}
{\Gamma^2(1+\sigma\lambda)\Gamma(\frac{1}{2}+\sigma(2+\lambda))} \\
\times &{}_5F_4
{\frac{1}{2}+\sigma\lambda,2\sigma(1+\lambda), \sigma(2+\lambda),
2\sigma,\sigma(1+\lambda) \choose
1+\sigma\lambda,1+2\sigma\lambda, \frac{1}{2}+\sigma(2+\lambda),
1+\sigma(1+\lambda)}.
\end{align}
\end{subequations}

\noindent Setting $\alpha=0$ in (\ref{alphazp1}) also recovers (\ref{azp1}) directly by Dixon's theorem.

\subsection{Radial sum rules and simplifications ($\alpha=0$)} \label{secsimp1}
\noindent  For $M>1$ the growth rate of the eigenvalues is sufficient to
ensure the convergence of the spectral determinants
\begin{equation}
D_-(E,M,\alpha,\lambda)\equiv D_-(0)\prod_{k=0}^\infty\left(1-\frac{E}{E_k^-}\right).
\end{equation}
\noindent Here $D_-(0)$ is a regularising prefactor designed to vanish when
some $E_k^-=0$. By analytic continuation we define
$D_+(E,\lambda)\equiv D_-(E,-\lambda)$ except for the values of
$\lambda$ given in (\ref{invalidlambda}).

\par In addition to the analytic continuation relating $D_-(E)$ and
$D_+(E)$, there is the quantum Wronskian equation
\begin{equation}  \label{qwradial}
2\lambda\omega^\frac{\alpha}{2}=
\overline{\omega}^{\lambda}D_-(\overline{\omega}E,-i\alpha)D_+(\omega E,i\alpha)
-\omega^\lambda D_-(\omega E,i\alpha)D_+(\overline{\omega} E,-i\alpha),
\end{equation}
where
\begin{equation}
\omega\equiv \exp(i\pi\sigma).
\end{equation}

\noindent The sets of eigenvalues $\{E_k^-\}$ and $\{E_k^+\}$,
having no relationship except at (\ref{invalidlambda}), are thus related to each
other by a simple expression involving their associated
spectral determinants.
\par Two clarifications need to be made regarding the quantum Wronskian
equation.
First is to note that when $\alpha\neq 0$ the spectral determinants
in (\ref{qwradial}) refer to
differential equations with imaginary coupling constants on the $x^{M-1}$
term, a problem which is remedied in \cite{Vorosex}. Secondly, for general
$\alpha$, the quantum Wronskian links together four different spectral
determinants, whereas setting $\alpha=0$ reduces this number to two.
For this latter case the subsequent results are much
less complicated
and so
$\alpha=0$ is fixed for the remainder of this section.

\par Using (\ref{qwradial}), we can derive sum rules relating together
different $Z_\mp(n)$. For small $E$ the spectral determinants
can be written in terms of their zeta functions \cite{Vorosspec0} as
\begin{equation} \label{dexpandzeta}
D_\mp(E)=D_\mp(0)\exp\left(-\sum_{n=1}^\infty\frac{Z_\mp(n)}{n}E^n\right).
\end{equation}

\noindent Using the technique of
\cite{Vorosquartic}, (\ref{dexpandzeta}) is substituted into (\ref{qwradial})
and the coefficients of the powers of $E$ are compared to obtain the radial sum
rules. The first few of these are
\begin{subequations} \label{rsr}
\begin{align}
    \label{rsr1} 0=N_{1}Z_-(1)+N_{-1}Z_+(1)&\\
    \label{rsr2}   0=N_{2}Z_-(2)+N_{-2}Z_+(2)
    &+\left(N^2_{1}-N_{2}\right)(Z_+(1)-Z_-(1))^2\\
    \nonumber
     0= N_{3}Z_-(3)+N_{-3}Z_+(3)
&-\frac{3}{2}\left(N_{3}-N_{2}N_{1}\right)(Z_+(2)-Z_-(2))(Z_+(1)-Z_-(1))\\
    \label{rsr3} &-\frac{1}{2}\left(N_{3}-3N_{2}N_{1}
    +2N^3_{1}\right)(Z_+(1)-Z_-(1))^3
\end{align}
\end{subequations}
\noindent where
\begin{equation} \nonumber
N_a\equiv \sin(\pi\sigma(\lambda+a))\csc(\pi\sigma\lambda).
\end{equation}

\noindent Under certain parameter choices, (\ref{rsr})
allows for $Z_+(n)$ to be given in terms of spectral zeta
functions with a simpler form. As an example, the
choice
$\sigma(\lambda+2)=m\in\mathbb{N}$ implies that $N_{2}=0$.
At such a point (\ref{rsr2}) does not include $Z_-(2)$
and hence, for $m\in\mathbb{N}$, we find
\begin{equation} \label{zp2simp}
Z_+\left(2,\sigma=m/(\lambda+2)\right)=-\frac{\sigma^{4-4\sigma}\Gamma^4(\sigma)\Gamma^2(1-2\sigma)\sin^3(2\pi\sigma)\csc^2(3\pi\sigma)}
{2^{4-4\sigma}\Gamma^2(1-3\sigma+m)\Gamma^2(1+\sigma-m)\sin(4\pi\sigma)}.
\end{equation}
\noindent This expression is much less complicated than the general form
for $Z_\mp(2)$ given in (\ref{azp2}),
reducing the ${}_5F_4$ hypergeometric series to a product of gamma and trigonometric functions.

\begin{exa} The anharmonic cubic oscillator has the zeta values
\begin{equation} \nonumber
Z_+(2)=\frac{8(\sqrt{5}-1)\pi^4}{5^\frac{17}{5}\Gamma^4(\frac{4}{5})
\Gamma^2(\frac{3}{5})}
\end{equation}
\noindent and
\begin{equation} \label{hypsimp}
 Z_-(3)=\frac{2^\frac{14}{5}3\pi^\frac{9}{2}\Gamma(\frac{7}{10})}
{5^\frac{23}{5}\Gamma^5(\frac{4}{5})\Gamma^2(\frac{9}{10})}-\frac{32\pi^6}
{5^\frac{51}{10}\Gamma^6(\frac{4}{5})\Gamma^3(\frac{3}{5})}
-\frac{2\pi\sqrt{5-2\sqrt{5}}}{5^\frac{21}{10}\Gamma(\frac{3}{5})}
\times {}_4F_3{\frac{3}{5},\frac{7}{10},\frac{4}{5},1 \choose \frac{7}{5},
\frac{3}{2},\frac{8}{5}}.
\end{equation}
\end{exa}
\vspace{2mm}
\noindent In addition to finding closed form expressions for $Z_\mp(n)$, there is a
reason to investigate the composite zeta functions
\begin{equation} \label{compositezeta}
Z(s)\equiv Z_+(s)+Z_-(s) \hspace{4mm} \textrm{and} \hspace{4mm} \tilde{Z}(s)\equiv Z_+(s)-Z_-(s).
\end{equation}
For the potential $V(x)=|x|^{2M}$ with eigenfunction
condition $\psi\in L^2(\mathbb{R})$ and eigenvalues $\{E_k\}$, $Z(s)$
and $\tilde{Z}(s)$ are respectively the `full' and `skew' zeta functions, defined
as
\begin{equation} \nonumber
Z(s)\equiv \sum_{k=0}^\infty \frac{1}{(E_k)^s} \hspace{4mm} \textrm{and} \hspace{4mm}
\tilde{Z}(s)\equiv \sum_{k=0}^\infty \frac{(-1)^k}{(E_k)^s}.
\end{equation}
\noindent For the anharmonic oscillators, these definitions correspond to (\ref{compositezeta}) as the spectrum $\{E_k\}$
is the interlacing union of $\{E_k^-\}$ and $\{E_k^+\}$. For general $\alpha$ and $\lambda$ this cannot be guaranteed.

\par Although generally used as a convenient notation, simplifications in the explicit forms of $Z(n)$ and $\tilde{Z}(n)$
will be of use later. The radial sum rules are used to find reduced forms for
$Z(2)$ and $\tilde{Z}(2)$ by realising points where $N_2=\pm N_{-2}$. For $m\in\mathbb{Z}^+$ we find
\begin{equation} \label{z2simp}
Z(2,\sigma=(2m-1)/2\lambda)=-\frac{\pi^4\sigma^{4-4\sigma}\Gamma^2(1-2\sigma)\sec^2(\pi\sigma)
\sec(2\pi\sigma)}
{4^{1-2\sigma}\Gamma^4(1-\sigma)\Gamma^2(\frac{3}{2}-\sigma-m)\Gamma^2
(\frac{1}{2}-\sigma+m)}
\end{equation}
and
\begin{equation} \label{tildez2simp}
\tilde{Z}(2,\sigma=1/4)=\frac{\pi^5\sec^2(\frac{\pi\lambda}{2})
\tan(\frac{\pi\lambda}{4})}{64\Gamma^4(\frac{3}{4})
\Gamma^2(\frac{3+\lambda}{4})\Gamma^2(\frac{3-\lambda}{4})}.
\end{equation}

\noindent There are no simplifications for $Z(2,\lambda=1/2)$
and no general simplifications for $\tilde{Z}(2)$, as $\lambda$
would conflict with (\ref{invalidlambda}).

\begin{exa} The anharmonic sextic oscillator has the zeta value
\begin{equation} \nonumber
\tilde{Z}(2)=\frac{(\sqrt{2}-1)\pi^5}{32\Gamma^4(\frac{3}{4})
\Gamma^2(\frac{7}{8})\Gamma^2(\frac{5}{8})}.
\end{equation}
\end{exa}

\subsection{Functional relations between hypergeometric series ($\alpha=0$)} \label{secspecderiv}

\noindent  While the radial sum rules in (\ref{rsr}) are useful for determining properties of $Z_\mp(n)$,
they can also be used to determine properties of the special functions
appearing in the explicit forms for $Z_\mp(n)$, as in (\ref{azp}).
The are two principle reasons that such derivations can be made. First is the
analytic continuation (\ref{ancon}) and the second is that the radial sum rules can be written as
\begin{subequations} \label{ps}
\begin{align}
\label{ps1} Z_+(1)&=-\frac{N_1}{N_{-1}}Z_-(1) \\
\label{ps2}  Z_+(2)&=
-\frac{N_2}{N_{-2}}Z_-(2)+
\left(\frac{N_2}{N_{-2}}-\frac{2N_1}{N_{-1}N_{-2}}+
\frac{N_1^2}{N_{-1}^2}\right)Z_-(1)^2.
\end{align}
\end{subequations}
These identities implicitly specify functional relations for the
 special functions appearing in (\ref{azp}). By substituting (\ref{azp1}
into (\ref{ps1}) and using (\ref{ancon}), we recover the identity
\begin{equation} \nonumber
\frac{\Gamma(\sigma(1-\lambda))}{\Gamma(1-\sigma(1+\lambda))}=
\frac{\Gamma(\sigma(1+\lambda))\sin(\pi\sigma(1+\lambda))}
{\Gamma(1-\sigma(1-\lambda))\sin(\pi\sigma(1-\lambda))},
\end{equation}
\noindent which is effectively two copies of the Euler reflection formula
for the gamma function.
\par The next application is to the zeta functions $Z_\mp(2)$, the
explicit forms involving a ${}_5F_4$ hypergeometric series as in (\ref{azp2}).
Therefore we expect to derive properties of the function
\begin{equation}{\cal F}(\lambda)\equiv
{}_5F_4
{\frac{1}{2}+\sigma\lambda,2\sigma(1+\lambda),\sigma(2+\lambda),
2\sigma,\sigma(1+\lambda) \choose
1+\sigma\lambda,1+2\sigma\lambda,\frac{1}{2}+\sigma(2+\lambda),
1+\sigma(1+\lambda)}.\end{equation}
\noindent
\noindent Substituting $Z_-(1)$ and $Z_\mp(2)$ into (\ref{ps2}) gives the functional relation
\begin{align}
  \label{5f4func}
& \frac{\sin(\pi\sigma(\lambda+2))4^{-\sigma\lambda}
\Gamma(2\sigma(1+\lambda))\Gamma(\sigma(2+\lambda))}
{\sin(\pi\sigma(\lambda-2))(1+\lambda)\Gamma^2(1+\sigma\lambda)
\Gamma(\frac{1}{2}+\sigma(2+\lambda))}{\cal F}(\lambda)=\\
 \nonumber &
\frac{4^{\sigma\lambda}\Gamma(2\sigma(1-\lambda))
\Gamma(\sigma(2-\lambda))}
{(\lambda-1)\Gamma^2(1-\sigma\lambda)\Gamma(\frac{1}{2}+\sigma(2-\lambda))}
{\cal F}(-\lambda)+
 \frac{4^{2\sigma-1}\sigma\pi^\frac{3}{2}\Gamma^2(1-2\sigma)
 \Gamma^2(\sigma(1+\lambda))}
{\Gamma^4(1-\sigma)\Gamma^2(1-\sigma(1-\lambda))
\Gamma(2\sigma)\sin^2(\pi\sigma)}
 \\
\nonumber & \hspace{25mm}
\times \left(\frac{\sin^2(\pi\sigma(1+\lambda))}
{\sin^2(\pi\sigma(1-\lambda))}-\frac{\sin(\pi\sigma(2+\lambda))}
{\sin(\pi\sigma(2-\lambda))}
-\frac{2\sin(\pi\sigma(1+\lambda))\sin(\pi\sigma\lambda)}
{\sin(\pi\sigma(1-\lambda))
\sin(\pi\sigma(2-\lambda))}\right).
\end{align}

\noindent Furthermore, numerical testing indicates that this identity appears to hold when
$0<\Re (\sigma)<3/4$, although our result only applies when
$\sigma\in\mathbb{R}$.
\par By choosing $\sigma$ and $\lambda$ to take special
values - precisely those which give simple forms for $Z_+(2)$ -
we recover further properties of ${\cal F}$.
For example the simplification (\ref{zp2simp}) gives, for $m\in\mathbb{N}$,
the identity
\begin{equation} \nonumber
{\cal F}\left(2-m/\sigma\right)
=\frac{(m-3\sigma)\Gamma^4(\sigma)\Gamma^2(1+2\sigma-m)\Gamma^2(1-2\sigma)
\Gamma(\frac{1}{2}+4\sigma-m)
\sin^3(2\pi\sigma)\csc^2(3\pi\sigma)}
{\sqrt{\pi}4^{m+1-4\sigma}
\Gamma^2(1-3\sigma+m)
\Gamma^2(1+\sigma-m)\Gamma(2\sigma)\Gamma(6\sigma-2m)\Gamma(4\sigma-m)
\sin(4\pi\sigma)}.
\end{equation}
\noindent When $m=1$
this identity is verified by Dixon's theorem but
we have been unable to find any known identities accounting for general
$m\in\mathbb{N}$. While ${\cal F}$ can be rewritten in terms of
${}_4F_3$ hypergeometric series \cite{Vorosquartic}, this appears to
provide no extra information and the functional identity (\ref{5f4func})
is not apparent.

\par The exact form of $Z_-(3)$ is known to contain Appell
series \cite{Vorosquartic}, functional relations for which being obtained by the above
methods. We would then expect relationships between two different
Appell series in terms of ${}_5F_4$ hypergeometric series.

\subsection{Functional relations between hypergeometric series ($\alpha\neq 0$)} \label{secspecderiv2}
The techniques used in Section \ref{secspecderiv} also apply
when $\alpha \neq 0$. In this situation
the sum rules found from (\ref{qwradial}) are more complicated,
featuring four separate zeta functions. Substituting (\ref{dexpandzeta})
into (\ref{qwradial}), the coefficient of $E$ in the subsequent expansion
gives, after the transformation $\alpha\rightarrow i \alpha$,
\begin{align} \nonumber
&D_-(0,\alpha)D_+(0,-\alpha)\left(\omega^{1-\lambda}Z_+(1,-\alpha)+\omega^{-(1+\lambda)}Z_-(1,\alpha)\right)
 \\ \label{alpharsr1}
& \hspace{20mm} =D_-(0,-\alpha)D_+(0,\alpha) \left(\omega^{\lambda-1}Z_+(1,\alpha)+\omega^{1+\lambda}Z_-(1,-\alpha)\right).
\end{align}

\noindent where $D_\mp(0)$ is given in \cite{Proof} as
\begin{equation} \label{d0}
D_-(0,\alpha) \propto (2\sigma)^{\frac{\alpha \sigma}{2}-\sigma\lambda-1/2}\frac{\Gamma(1+2\sigma\lambda)}{\Gamma(\frac{1}{2}+\frac{\alpha \sigma}{2}+\sigma\lambda)}.
\end{equation}

\noindent The explicit form for $Z_-(1)$ features a ${}_3F_2$ hypergeometric series as in (\ref{alphazp1}). Therefore
(\ref{alpharsr1}) is implicitly a four-term functional relation involving this function. Splitting into real
and imaginary parts, (\ref{alpharsr1}) is actually two independent functional relations
satisfied by the ${}_3F_2$ hypergeometric series. We define the function
\begin{equation}
\label{calgdef}
{\cal G}(\alpha,\lambda)\equiv \frac{1}{\Gamma(\frac{1}{2}+2\sigma+\frac{\sigma\alpha}{2}+\sigma\lambda)\Gamma(\frac{1}{2}-\frac{\sigma\alpha}{2}-\sigma\lambda)}
{}_3F_2
{\frac{1}{2}+\frac{\sigma\alpha}{2}+\sigma\lambda,2\sigma(1+\lambda),
2\sigma \choose 1+2\sigma\lambda,\frac{1}{2}+\frac{\sigma\alpha}{2}+\sigma(2+\lambda)},
\end{equation}

\noindent where the hypergeometric series converges for $\Re(\sigma)<1/2$. Substituting (\ref{alphazp1})
and (\ref{d0}) into (\ref{alpharsr1}) gives, after taking real and imaginary parts,
the pair of four-term relations

\begin{subequations} \label{calgident}
\begin{equation} \label{calgident1}
{\cal G}(\alpha,\lambda)+{\cal G}(-\alpha,\lambda)
=
\frac{\sin(\pi\sigma(1-\lambda))\Gamma(1+2\sigma\lambda)\Gamma(2\sigma(1-\lambda))}{\sin(\pi\sigma(1+\lambda))\Gamma(1-2\sigma\lambda)\Gamma(2\sigma(1+\lambda))}
\left({\cal G}(\alpha,-\lambda)+{\cal G}(-\alpha,-\lambda)\right)
\end{equation}
\noindent and
\begin{equation} \label{calgident2}
{\cal G}(\alpha,\lambda)-{\cal G}(-\alpha,\lambda)
=
\frac{\cos(\pi\sigma(1-\lambda))\Gamma(1+2\sigma\lambda)\Gamma(2\sigma(1-\lambda))}{\cos(\pi\sigma(1+\lambda))\Gamma(1-2\sigma\lambda)\Gamma(2\sigma(1+\lambda))}
\left({\cal G}(\alpha,-\lambda)-{\cal G}(-\alpha,-\lambda)\right).
\end{equation}
\end{subequations}

\noindent This pair of equations allows for the derivation of three-term
equations featuring $G(\alpha,\lambda)$. One such example is given
by

\begin{equation} \label{gthreeterm}
{\cal G}(\alpha,\lambda)=\frac{\Gamma(1+2\sigma\lambda)\Gamma(2\sigma(1-\lambda))\Gamma(1-2\sigma(1+\lambda))}{\pi\Gamma(1-2\sigma\lambda)}
\left(\sin(2\pi\sigma){\cal G}(\alpha,-\lambda)-\sin(2\pi\sigma\lambda){\cal G}(-\alpha,-\lambda)\right).
\end{equation}

\noindent Now ${\cal G}(\alpha,\lambda)$ has been shown to satisfy a particular three term
functional relation. From (\ref{gthreeterm}), specific parameter choices allow for
identities which only feature ${\cal G}(\alpha,\lambda)$ twice. One example
is given by $\lambda=1-1/2\sigma$. When this is the case
${\cal G}(\alpha,\lambda)$ simplifies to a
${}_2F_1$ hypergeometric series, which is in-turn expressed in terms of gamma functions
by Gauss' theorem. The subsequent relation is that

\begin{equation}
\frac{\Gamma^2(1-2\sigma)\Gamma(2-2\sigma)}{\Gamma(2\sigma)\Gamma(1-\sigma(1+\delta))\Gamma(1-\sigma(1-\delta))\Gamma(2-4\sigma)}
={\cal G}(2\delta,1-1/2\sigma)+{\cal G}(-2\delta,1-1/2\sigma).
\end{equation}

\noindent Similar such reductions can be found by any simplification of the
${}_3F_2$ hypergeometric series in (\ref{calgdef}) being substituted into
the three-term relation in (\ref{gthreeterm}).

\section{${\cal PT}$-symmetric Schr\"odinger problems} \label{secpt}
\begin{figure*}[t]
\begin{center}
\caption{Stokes rays for $M=2.7$ and $K=1,2,3$ are given by the solid (blue) lines.
The dashed (red) lines are the relevant anti-Stokes rays and we choose
the quantisation contour to asymptotically tend towards these,
while avoiding the branch cut along the imaginary axis. The different boundary
conditions thus define separate spectra.}
\label{stokespic1}
\includegraphics[width=0.32\textwidth]{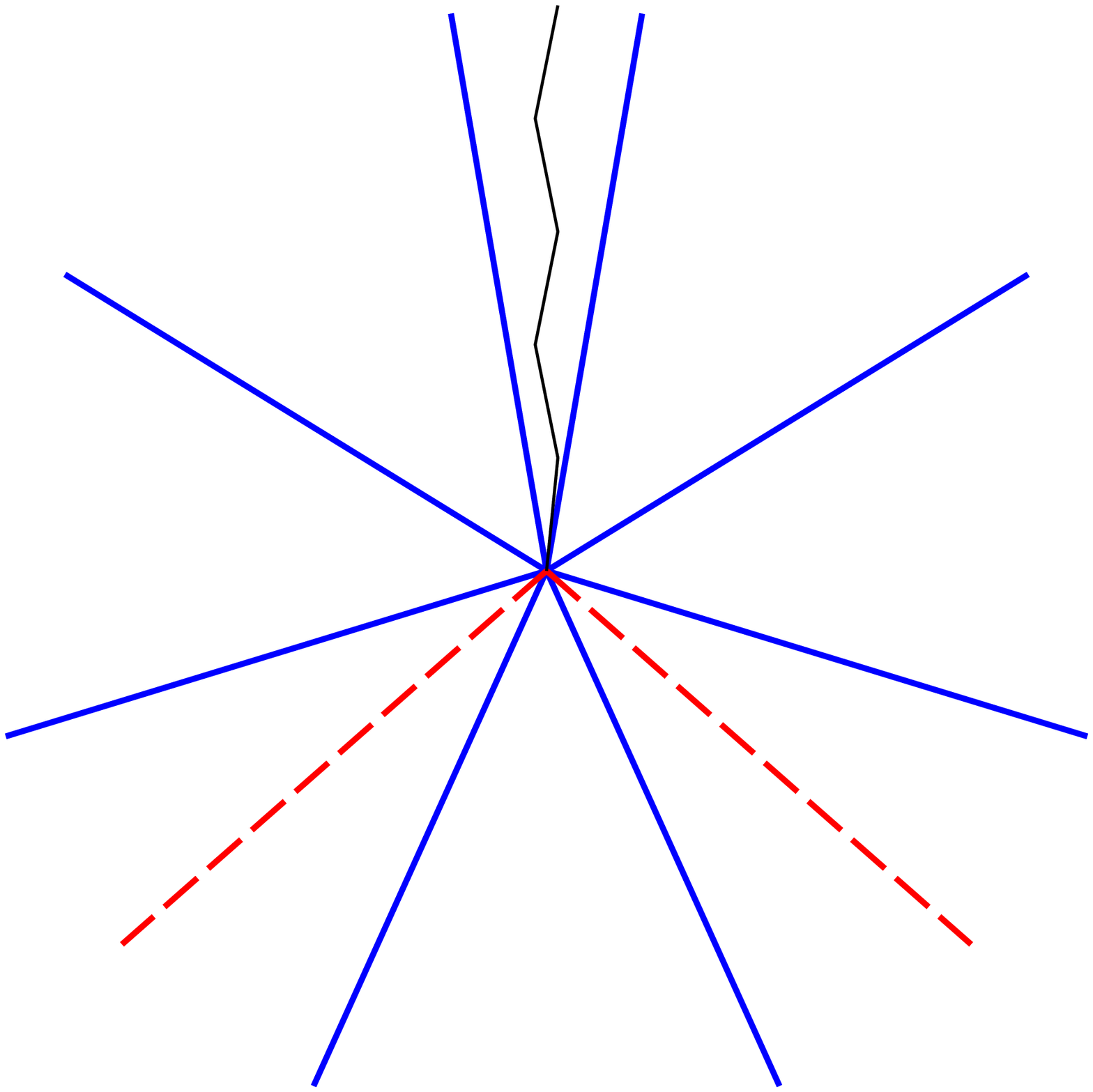}
\includegraphics[width=0.32\textwidth]{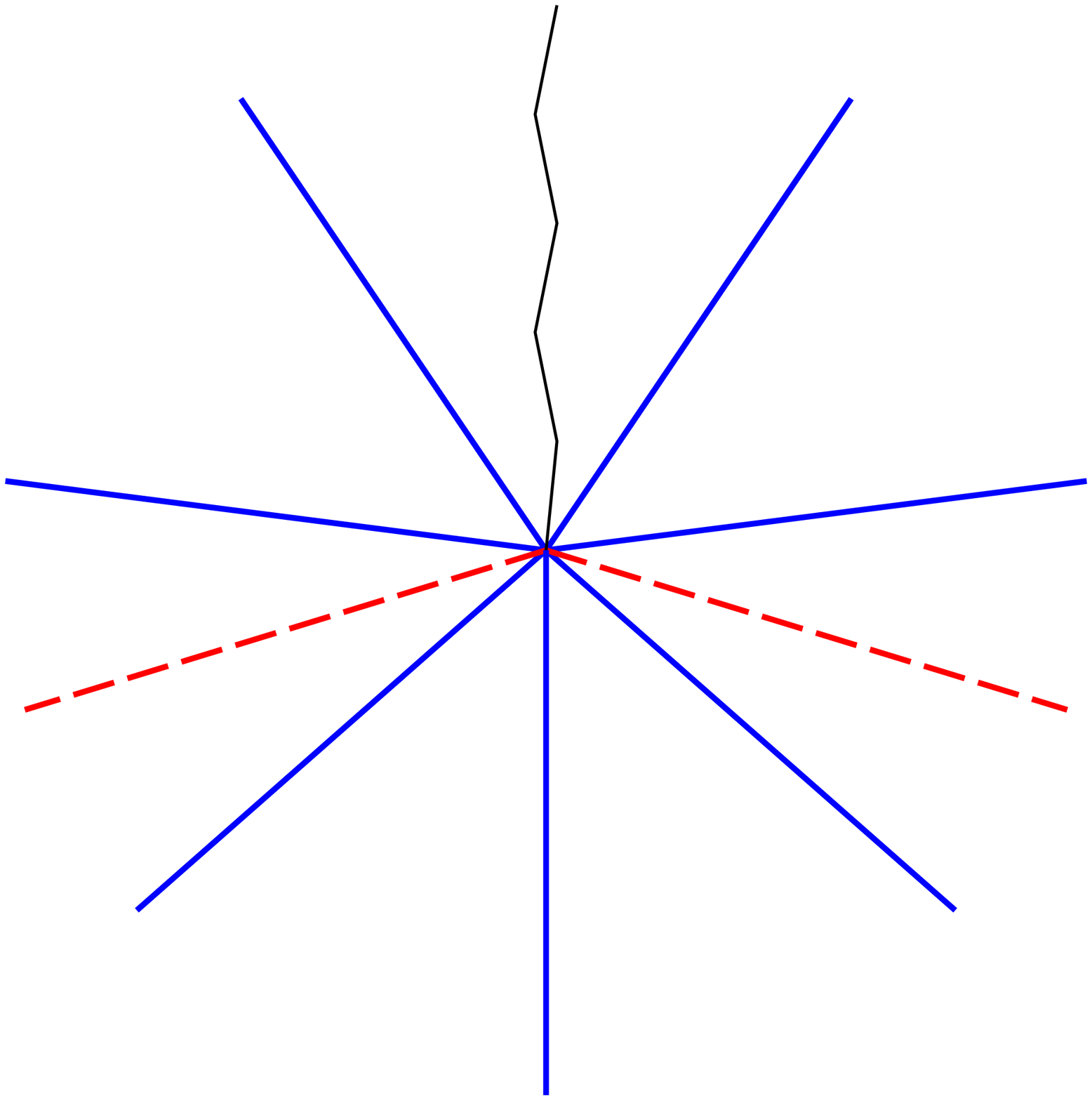}
\includegraphics[width=0.32\textwidth]{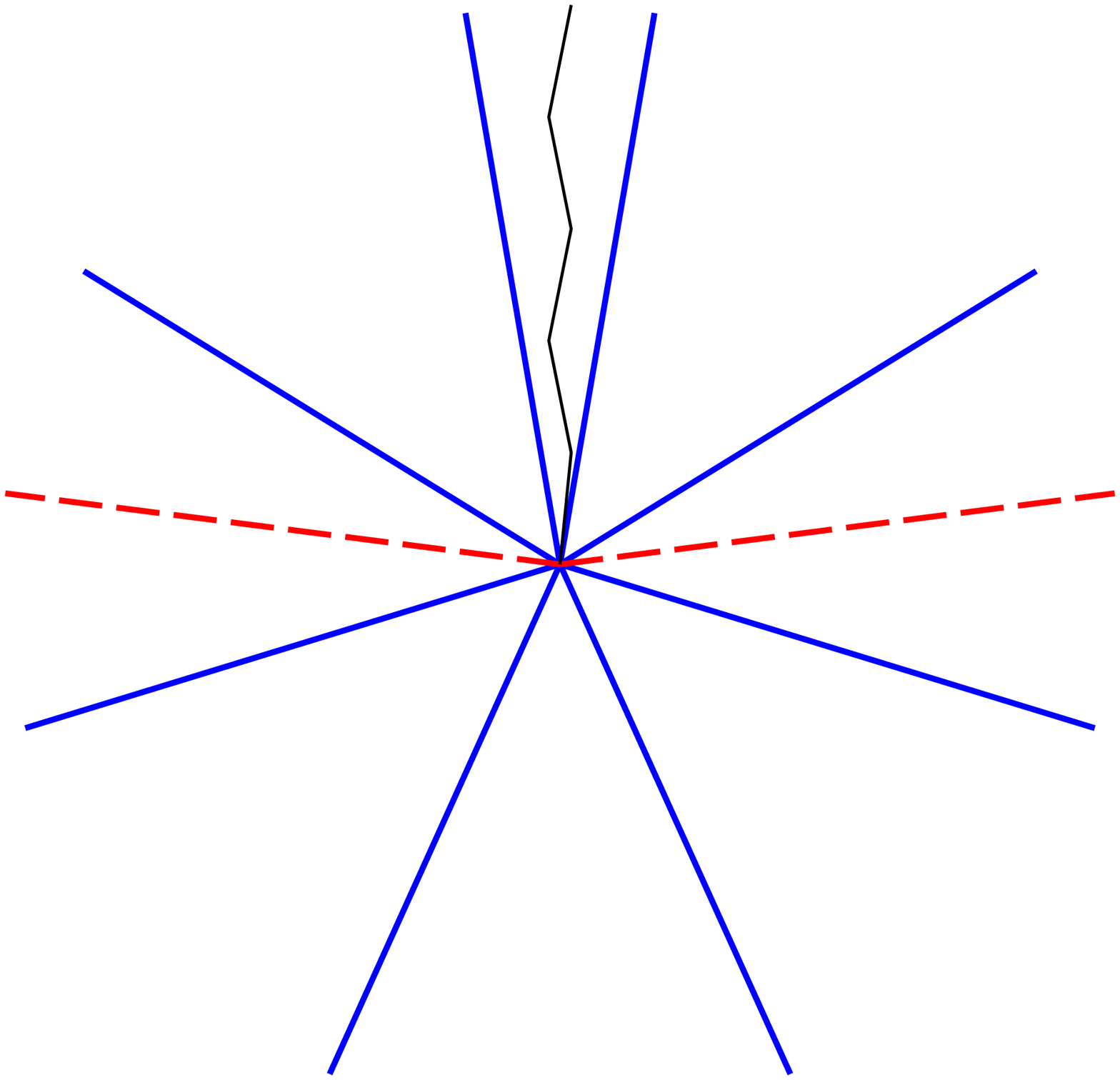}
\end{center}
\end{figure*}

Now we consider a class of non-Hermitian eigenvalue problems related to the radial problems
of the previous section, given for $K\in\mathbb{N}$ by the Schr\"odinger equation
\begin{equation} \label{kproblems}
-\phi''+\left((-1)^{K}(ix)^{2M}-\alpha(ix)^{M-1}+
\frac{\lambda^2-\frac{1}{4}}{x^2}\right)\phi=E\phi.
\end{equation}
\noindent The boundary conditions are that
$\phi\in L^2({\cal C}(x))$ where ${\cal C}(x)$ is a quantisation contour that asymptotes
towards the anti-Stokes rays with complex arguments
$-\pi/2\pm\pi(K+1)/(2M+2)$. These anti-Stokes rays are shown in Figure \ref{stokespic1}. Such a contour must also avoid the branch cut
along the positive imaginary axis, employed to ensure that the potential is
single-valued for general $M$.
\par Surprisingly these Schr\"odinger problems,
with complex potentials and boundary conditions defined in the complex plane,
can have an entirely real spectrum \cite{Bender, Wronskian, Proof, Shin}.
The suggested reason for the possibility of spectral reality lies in the ${\cal PT}$-symmetry
of the problem, meaning that (\ref{kproblems}) and the boundary conditions are invariant under a reflection in the imaginary axis.
Specifically the spectrum is known to be real if the ${\cal PT}$-symmetry is unbroken, meaning that the eigenfunctions are invariant under the combined ${\cal PT}$ operation \cite{Wronskian}. The revelation
of a non-Hermitian problem exhibiting real eigenvalues has since spawned
a huge research effort with hundreds of papers. Recent review articles of the
subject are also available \cite{Makingsense, Review}.

\par Rather than investigating spectral reality, our interest
will be in the relationships between the radial problems in the previous section
and the ${\cal PT}$-symmetric of this section. For fixed $M$, different choices of $K$ will result
in quantisation contours which asymptotically lie within different pairs of Stokes sectors. Generally these different problems are then expected
to have unique spectra for fixed $M,\alpha,\lambda$, as was demonstrated in
\cite{Wronskian}. To these separate spectra we associate the eigenvalues
$\{E_k^K\}$ and the spectral functions
\begin{equation}C_K(E)\equiv C_K(0)\prod_{k=0}^\infty\left(1-\frac{E}{E_k^K}\right)
\hspace{6mm} \textrm{and} \hspace{6mm}
{\cal Z}_K(s)\equiv \sum_{k=0}^\infty \frac{1}{(E_k^K)^s}.\end{equation}
\noindent Despite the different spectra appearing to have no obvious analytic relationship, in \cite{Sibuya,On} a number of results are given which link together
different $C_K(E)$ and hence the associated eigenvalues. Following work in \cite{On}, using the normalisations
in \cite{Draft}, one such result is the `fused' quantum Wronskian equation
\begin{align}
  \nonumber
2\lambda\omega^{\frac{\alpha}{2}(1+(-1)^KK)}
C_K(-E,\alpha)&=
\overline{\omega}^{(K+1)\lambda} D_-(\overline{\omega}^{K+1}E,(-i)^{K+1}\alpha)
D_+(\omega^{K+1} E,i^{K+1}\alpha) \\
\label{qw}
&
-\omega^{(K+1)\lambda} D_-(\omega^{K+1}E,i^{K+1}\alpha)
D_+(\overline{\omega}^{K+1}E,(-i)^{K+1}\alpha),
\end{align}
\noindent where $M>1$ and $\lambda$ is again restricted to exclude the values in
(\ref{invalidlambda}).
\par The fused quantum Wronskian is essentially a simple
extension of (\ref{qwradial}), which can be verified
by setting $K=0$ and $C_0(E)\equiv 1$ in (\ref{qw}). The function
$C_0(E)$ is understood to be a constant as the choice $K=0$ dictates that the quantisation contour
of the eigenvalue problem asymptotically lies within two adjacent Stokes sectors, meaning
an empty spectrum \cite{On}. Therefore
any sum rules derived from (\ref{qw}) must reproduce those
derived from (\ref{qwradial}) by setting $K=0$
and defining $Z_0(s)\equiv 0$.
\par In contrast to (\ref{qwradial}), the fused quantum Wronskian
will feature only two spectral determinants on the right hand side
when $\alpha \neq 0$ and $K$ is odd. At such values of $K$ the sum rules must
be of similar form to those in (\ref{rsr}).

\subsection{Fused sum rules and simplifications} \label{secfused}
Calculating ${\cal Z}_K(n)$ directly can be handled by the
Green's function method used previously. This was
first implemented in \cite{Zeta,Yes} for the
 anharmonic ($\alpha=0$, $\lambda=1/2$) oscillators, where the results were used to verify
conjectures on the spectral reality of the ${\cal PT}$-symmetric
problems. However there is an easier method than direct calculation,
using sum rules and the exact forms for $Z_\mp(n)$ calculated in (\ref{alphazp}) and (\ref{azp}). To determine sum rules between ${\cal Z}_K(n)$ and $Z_\mp(n)$,
$C_K(-E)$ is expanded near the origin as
\begin{equation} \label{ckexpandzeta}
C_K(-E)=C_K(0)\exp\left(-\sum_{n=1}^\infty\frac{{\cal Z}_K(n)}{n}
(-E)^n\right).
\end{equation}
\noindent The fused sum rules are calculated by substituting (\ref{dexpandzeta}) and (\ref{ckexpandzeta}) into (\ref{qw})
and comparing coefficients of the powers of $E$. Temporarily suppressing the dependence on $K$ when $\alpha\neq 0$, the first few fused sum rules are
\begin{subequations} \label{fsr}
\begin{align}
     \label{fsr1}  {\cal Z}_K(1)&=-L_{1}Z_-(1)-L_{-1}Z_+(1)\\
     \label{fsr2} {\cal Z}_K(2)&=L_{2}Z_-(2)+L_{-2}Z_+(2)  +\left(L^2_{1}-L_{2}\right)
    \tilde{Z}(1)^2\\
    \nonumber
    {\cal Z}_K(3)&=-L_{3}Z_-(3)-L_{-3}Z_+(3)+\frac{3}{2}\left(L_{3}-L_{2}L_{1}\right)\tilde{Z}(2)\tilde{Z}(1)\\
    \label{fsr3} &\hspace{43.5mm}+\frac{1}{2}\left(L_{3}-3L_{2}L_{1}+2L^3_{1}\right)\tilde{Z}(1)^3
\end{align}
\end{subequations}
\noindent where
\begin{equation} \nonumber
L_a\equiv \sin(\pi\sigma(K+1)(\lambda+a))\csc(\pi\sigma(K+1)
\lambda).
\end{equation}
\noindent For $\alpha=0$ the fused sum rules apply for all $K$, whereas
for $\alpha \neq 0$ this is only true when $K$ is odd. In the latter
situation ${\cal Z}_K(n,\alpha)$ is given in terms of $Z_\mp(n,-\alpha)$
for $K=4n-3$ and in terms of $Z_\mp(n,\alpha)$ for $K=4n-1$.

\par To calculate ${\cal Z}_K(n)$ it is now
only necessary to calculate $Z_-(n),\dots,Z_-(1)$, determine
$Z_+(n,\lambda)$ from (\ref{ancon}) and then substitute these into the
appropriate fused sum rule. This method implies that ${\cal Z}_K(n)$ is always
real-valued, despite no guarantee that the eigenvalues will be real. The explanation
is that all ${\cal PT}$-symmetric systems have a real characteristic
equation, implying that the eigenvalues are either real or appear in complex conjugate pairs \cite{Realchar}.

\par Given the fundamentally differences between the radial problems and the ${\cal PT}$-symmetric problems,
there is a priori no reason to expect such simple connections between the
differing sets of spectra. Specific examples of the fused sum rules can be written in many ways, depending on the relationships between $Z_-(n)$ and $Z_+(n)$
as determined by the radial sum rules (\ref{rsr}).

\begin{exa} The anharmonic quartic oscillator has the
sum rules
\begin{subequations}
\begin{align}
\nonumber {\cal Z}_1(1)&=2Z_-(1)\\
 \nonumber {\cal Z}_1(2)&=Z_-(1)^2-Z_-(2)\\
\nonumber 2{\cal Z}_1(3)&=3Z_-(2)Z_-(1)+Z_-(1)^3.
\end{align}
\end{subequations}
\end{exa}

\noindent As with the radial sum rules, the fused sum rules are also useful for deriving
simplifications of ${\cal Z}_K(n)$. For example, comparing (\ref{rsr2})
and (\ref{fsr2}) shows for $m\in\mathbb{N}$ that
\begin{align}
 \nonumber &{\cal Z}_K(2,\sigma=m/(\lambda+2),\alpha=0)=
\frac{\pi^4\sigma^{4-4\sigma}\Gamma^2(1-2\sigma)\csc^2(\pi\sigma)}
{16^{1-\sigma}\Gamma^2(1+\sigma-m)\Gamma^2(1-3\sigma+m)\Gamma^4(1-\sigma)}\times\\
\label{genreduc}  & \left(\frac{(\csc(3\pi\sigma)+\csc(\pi\sigma))^2}
{\csc^2(\pi\sigma(K+1))\sin^2(2\pi\sigma(K+1))}
-\frac{\sin(4\pi\sigma(K+1))(1+\sec(2\pi\sigma))}
{(1+2\cos(2\pi\sigma))^2 \sin^2(\pi\sigma)\sin(2\pi\sigma(K+1))}\right).
\end{align}

\begin{exa} The anharmonic cubic oscillator
has the zeta value
\begin{equation} \nonumber
{\cal Z}_1(2)=\left(1-\frac{2}{\sqrt{5}}\right)Z_+(1)^2=
\frac{16(\sqrt{5}-2)\pi^4}
{5^\frac{29}{10}\Gamma^4(\frac{4}{5})\Gamma^2(\frac{3}{5})}.
\end{equation}
\end{exa}

\noindent Certain choices of $M$ and $K$ reduce the
quantum Wronskian to the form $C_K(-E)\propto D_+(\pm E)D_-(\pm E)$ and hence
give the zeta function identity
${\cal Z}_K(n)=(\mp 1)^n Z(n)$. Additionally (\ref{fsr})
will allow for the form ${\cal Z}_K(n)=\tilde{Z}(n)+f(Z_\mp(n-1),\dots,Z_\mp(1))$
under appropriate parameter choices. Therefore the
simplifications for $Z(2)$ and $\tilde{Z}(2)$, as in (\ref{z2simp})
and (\ref{tildez2simp}), apply to ${\cal Z}_K(2)$.
Two examples of these are :

\begin{exa} The
quartic oscillator has the zeta value
\begin{equation} \nonumber
{\cal Z}_2(2,\alpha=0,\lambda=3/2)=
\left(\frac{3}{2}\right)^\frac{1}{3}\Gamma^2\left(\frac{2}{3}\right).
\end{equation}
\end{exa}

\begin{exa} The anharmonic sextic oscillator has the zeta value
\begin{equation} \nonumber
{\cal Z}_2(2)=\frac{(3-2\sqrt{2})\pi^5}{16\Gamma^4(\frac{3}{4})
\Gamma^2(\frac{7}{8})\Gamma^2(\frac{5}{8})}.
\end{equation}
\end{exa}

\subsection{Nonlocal integrals of motion} \label{secim}
An application for the fused sum rules in (\ref{fsr}) is within the `ODE/IM Correspondence'
\cite{Anharmonic,BLZSpec,On,Alpha2} which links
the eigenvalue problems in this paper to a class
 of integrable models. The correspondence, which was established
 by functional relations such as the quantum Wronskian,
 is summarised in-depth in the review article \cite{Review}.

 \par Two essential components for realising the correspondence are
 the continuum analogues of the $\mathbb{T}$- and $\mathbb{Q}$-operators
 which appear in the study of integrable systems such as the six-vertex
 model. These analogues were introduced in \cite{BLZ1,BLZ2} to study
 the integrable structure of conformal field theory. When $M>1$ and $\alpha=0$ the vacuum eigenvalues of these
analogues, denoted $T(s)$ and $Q(s)$, are related simply to the spectral determinants of the radial and ${\cal PT}$-symmetric problems. This remarkable connection was shown in \cite{BLZSpec,On} and is summarised by the relations
\begin{equation} \label{thiseqn22}
T(s)=C_1(-\nu s^2)
\end{equation}
\noindent and
\begin{equation} \label{thiseqn23}
Q(s)=\frac{1}{D_-(0)}D_-(\nu s^2)
\end{equation}
\noindent under the parameter identifications
\begin{equation} \label{varsident}
\beta^2=\sigma, \hspace{4mm} p=\frac{\sigma\lambda}{2}
\hspace{4mm} \textrm{and} \hspace{4mm}
\nu\equiv \left(\frac{\sigma}{2}\right)
^{2\sigma-2}\Gamma^2(1-\sigma).
\end{equation}

\noindent Considering (\ref{thiseqn23}), direct relations are given
in \cite{BLZ2} between the nonlocal integrals of motion (IMs) $H_n$, used in a power
series expansion for $Q(s)$, and the zeta functions $Z_-(n)$. Additionally
conjectured in \cite{BLZ2} were analytic continuations for $Z_-((2n-1)/(2\sigma-2))$ in
terms of local IMs $I_{2n-1}$ and for $Z_-(-n/\sigma)$ in
terms of nonlocal IMs $\tilde{H}_n$. Therefore there appear to be good reasons to consider the spectral zeta functions in relation to the nonlocal IMs. Specifically we will work with $G_n$, which
were given in \cite{BLZ1} to be nonlocal IMs appearing in the power series expansion
\begin{equation} \label{texp}
T(s)=T(0)+\sum_{n=1}^\infty G_n s^{2n}.
\end{equation}
\noindent These functions are found directly from the integral
\begin{align}
\nonumber   G_n\equiv
2\int^{2\pi}_0du_1\int^{u_1}_0dv_1\int^{v_1}_0du_2\int^{u_2}_0dv_2
\dots \int^{v_{n-1}}_0du_n \int^{u_n}_0dv_n \cos \left(2p\left( \pi+ \sum_{i=1}^n v_i-u_i\right)\right) \\
\label{gint}
\prod_{j>i}^n\left[4\sin\left(\frac{u_i-u_j}{2}\right)
\sin\left(\frac{v_i-v_j}{2}\right)\right]^{2\beta^2} \prod_{j>i}^n\left[2\sin\left(\frac{v_i-u_j}{2}\right)\right]^{-2\beta^2}
\prod_{j\geq i}^n\left[2\sin\left(\frac{u_i-v_j}{2}\right)\right]^{-2\beta^2},
\end{align}
\noindent with the first of these being given simply as
\begin{equation} \label{g11}
G_1=\frac{4\pi^2\Gamma(1-2\beta^2)}
{\Gamma(1-\beta^2-2p)\Gamma(1-\beta^2+2p)}.
\end{equation}

\noindent The nested integration required to calculate $G_n$ is reminiscent of the nested integration required to compute
$Z_-(n)$ as in Section \ref{seccomp}. The difference is that
while $Z_-(2)$ can be written exactly, no general form
for $G_2$ is known, although
alternative approaches have been used in \cite{BLZKondo,BLZExcited,BLZNon}.
To calculate $G_n$, (\ref{thiseqn22}) is expanded on both sides as a power series by using
(\ref{ckexpandzeta}) and (\ref{texp}). Comparing powers of $s$
and using
$T(0)=C_1(0)=2\cos(\pi\sigma\lambda)$ \cite{BLZ1,On}, the first few $G_n$ are
found:
\begin{subequations} \label{g}
\begin{align}
\label{g1}  G_1&=2\left(\frac{\sigma}{2}\right)^{2\sigma-2}\Gamma^2(1-\sigma)
\cos(\pi\sigma\lambda){\cal Z}_1(1)\\
\label{g2}  G_2&=\left(\frac{\sigma}{2}\right)^{4\sigma-4}\Gamma^4(1-\sigma)
\cos(\pi\sigma\lambda)\left({\cal Z}_1(1)^2-{\cal Z}_1(2)\right) \\
\label{g3}  G_3&=
\frac{1}{6}\left(\frac{\sigma}{2}\right)^{6\sigma-6}\Gamma^4(1-\sigma)
\cos(\pi\sigma\lambda)\left({\cal Z}_1(1)^3+2{\cal Z}_1(3)-3{\cal Z}_1(2)
{\cal Z}_1(1)\right)
\end{align}
\end{subequations}
\noindent Now that $G_n$ are expressed in terms of the zeta functions ${\cal Z}_1(n)$, they can be written in terms of $Z_\mp(n)$ by (\ref{rsr}). This circumvents the need
to calculate $G_n$ directly, expressing them instead in terms
of zeta functions which can be calculated from work in the previous sections. Given that $Z_-(n)$ becomes more complicated as $n$ increases, identities for $G_n$ are expected to reflect this property. However the integral expression for $Z_-(n)$ in (\ref{gennested}) can always be given in terms of some infinite power series. This provides a computational advantage over (\ref{gint}), which has not been directly evaluated for $n\geq 2$.
\par The validity of (\ref{g}) can be checked by first substituting (\ref{azp1}) into
(\ref{fsr1}). The general identity for ${\cal Z}_1(1)$ is then
substituted into (\ref{g1}) and, after the variable changes in
(\ref{varsident}), found to agree with (\ref{g11}) exactly.
The process is simply extended to $G_n$ and the higher nonlocal IMs
giving, as an example, that
\begin{align}
 \nonumber  G_2=\frac{4\pi^4\Gamma^2(1-2\sigma)\sec(\pi\sigma\lambda)}
{\Gamma^2(1-\sigma(1-\lambda))\Gamma^2(1-\sigma(1+\lambda))}
\left(1-\frac{\cos^4(\pi\sigma)}{\sin^2(\pi\sigma(1-\lambda))
\sin^2(\pi\sigma(1+\lambda))}\right)
\\ \label{g2}
+ \left(\frac{\sigma}{2}\right)^{4\sigma-4}\Gamma^4(1-\sigma)
\cos(\pi\sigma\lambda)\left(\frac{\sin(2\pi\sigma(2-\lambda))}
{\sin(2\pi\sigma\lambda)}Z_+(2)
- \frac{\sin(2\pi\sigma(2+\lambda))}{\sin(2\pi\sigma\lambda)}Z_-(2)\right)
\end{align}
\noindent where the identifications in (\ref{varsident}) must be taken into account.

This expression for $G_2$ can be checked to agree with some special
cases given in \cite{BLZExcited}. Practically (\ref{g2}) provides little
extra information over the methods given in \cite{BLZKondo,BLZExcited,BLZNon}
due to the lack of known analytic properties of the ${}_5F_4$ hypergeometric
terms appearing in (\ref{azp2}). However the simplifications for ${\cal Z}_K(2)$,
for example (\ref{genreduc}), allow for concise special values of $G_2$ to be found, in some cases being as simple as the general formula
for $G_1$.

\begin{exa} For $\beta^2=2/5$ and $p=11/10$ the second nonlocal IM is given by
\begin{equation} \nonumber
G_2=\frac{32(5+\sqrt{5})\pi^4\Gamma^2(\frac{3}{5})}{45\Gamma^4(\frac{4}{5})}.
\end{equation}
\end{exa}

\noindent The above methods can be extended to compute $G_3$ and higher nonlocal
IMs. Simplifications available by the sum rules would mean that, for specific
parameter choices, $G_3$ can be given without having to calculate a general
form for $Z_-(3)$ directly as in Section \ref{seccomp}.

\section{Summary and future work}
We have calculated the spectral zeta functions $Z_\mp(1)$ and $Z_\mp(2)$
pertaining to the radial problems in (\ref{ode}). Using the quantum Wronskian, sum rules were established which give relationships
between the different $Z_\mp(n)$. As well as illustrating the deep connections between the two problems, these sum rules allow for the simplification of $Z_\mp(2)$ with many examples found in the anharmonic oscillators. The physical reasons for such simplifications, if any, is not yet understood.
\par Our methods were then applied to determine special properties of the gamma function and specific hypergeometric series. We have not found another way to obtain the functional relations between these hypergeometric series and a separate derivation would be of interest. This technique could be applied in general to any Schr\"odinger problems where sum rules can be established between
the zeta functions. Furthermore the functional relations established
 encodes many known, closed-form identities for specific hypergeometric series. A full study of this topic may therefore prove interesting in the future.
\par In Section 3 the zeta functions ${\cal Z}_K(n)$ of a related, ${\cal PT}$-symmetric problem were calculated not directly but using the fused sum rules relating them to $Z_\mp(n)$. This requires fewer calculations than the direct method and serves to highlight the deep connections between the real-line and monodromy problems. After calculating ${\cal Z}_K(n)$, we showed how the ODE / IM correspondence can be used to calculate the vacuum nonlocal IMs $G_n$.
This was restricted to $\alpha=0$, although we expect our work will be useful in testing the correspondence postulated in \cite{Alpha2,Alpha3,Alpha4}, where this restriction is not imposed.

\par We deliberately excluded the possibility of a zero-energy eigenvalue.
However our initial investigations show that this need not be the case.
In particular we have found that spectral zeta functions can be used to
detect the presence (and classify the order) of zero-energy exceptional
points in non-Hermitian eigenvalue problems. A natural application of
this method is
to the ${\cal PT}$-symmetric problems specified in (\ref{kproblems}),
where the fused sum rules are used. We hope to work more on this in the
future.
\newline

\noindent \textbf{Acknowledgements}
\newline \noindent
I am very grateful to my supervisor
Clare Dunning for many useful ideas and suggestions. Additionally I would
like to thank Andr\'e Voros for helpful conversations and correspondence
and Paulo Assis for generously giving his time for discussion. Finally I wish
to thank Andy Hone for formatting and presentation advice.
My studies are funded jointly by EPSRC (EP/P502543/1) and the University of
Kent.


\begin{thebibliography}{99}
\bibitem{Alpha3}
P. E. G. Assis, Quantum Physics with Non-Hermitian Operators, Conference talk, Dresden (2011)
\bibitem{Alpha4}
P. E. G. Assis, Private communication (2011)
\bibitem{Baxter}
R. J. Baxter, Exactly Solved Models in Statistical Mechanics,(New York: Academic 1982)
\bibitem{BLZ1}
V. V. Bazhanov, S. L. Lukyanov and A. B. Zamolodchikov, Integrable structure of conformal field theory. Quantum KdV theory and
Thermodynamic Bethe Ansatz, \textit{Commun. Math. Phys}. \textbf{17} (1996) 381-398
\bibitem{BLZ2}
V. V. Bazhanov, S. L. Lukyanov and A. B. Zamolodchikov, Integrable structure of conformal field theory II. Q-operator and DDV equation, \textit{Commun. Math. Phys.} \textbf{190} (1996) 247-278
\bibitem{BLZExcited}
V. V. Bazhanov, S. L. Lukyanov and A. B. Zamolodchikov, Integrable quantum field theories in finite volume: excited state energies,
\textit{Nucl. Phys. B} \textbf{489} (1997) 487-531
\bibitem{BLZNon}
V. V. Bazhanov, S. L. Lukyanov and A. B. Zamolodchikov, On non-equilibrium states in QFT model with boundary interaction,
\textit{Nucl. Phys. B} \textbf{549} (1999) 529-545
\bibitem{BLZSpec}
V. V. Bazhanov, S. L. Lukyanov and A. B. Zamolodchikov, Spectral determinants for Schr\"odinger equation and Q-operators of conformal field theory,
\textit{J. Stat. Phys.} \textbf{102} (2001) 567-576
\bibitem{Alpha2}
V. V. Bazhanov and Z. Tsuboi, Baxter's Q-operators for supersymmetric spin chains, \textit{Nucl. Phys. B} \textbf{805} (2008) 451-516
\bibitem{Makingsense}
C. M. Bender, Making sense of non-Hermitian Hamiltonians, \textit{Rept. Prog. Phys.} \textbf{70} (2007) 947
\bibitem{Realchar}
C. M. Bender, M. V. Berry and A. Mandilara, Generalized ${\cal PT}$-symmetry and real spectra, \textit{J. Phys. A} \textbf{35} (2002) L467-471
\bibitem{Bender}
C. M. Bender and S. Boettcher, Real spectra in non-Hermitian Hamiltonians having ${\cal PT}$-symmetry, Phys. Rev. Lett. \textbf{80} (1998) 5243-5246
\bibitem{Wronskian}
C. M. Bender, S. Boettcher and P. N. Meisinger, ${\cal PT}$-symmetric quantum mechanics, \textit{J. Math. Phys}. \textbf{40} (1999) 2201-2230
\bibitem{QES2}
C. M. Bender and G. V. Dunne, Quasi-exactly solvable systems and orthogonal polynomials, \textit{J. Math. Phys.} \textbf{37} (1996) 6-11
\bibitem{Yes}
C. M. Bender and Q. Wang, Comment on a recent paper by Mezincescu, \textit{J. Phys. A} \textbf{34} (2001) 3325-3328.
\bibitem{Crandall}
R. E. Crandall, On the quantum zeta function,  \textit{\textit{J. Phys. A}} \textbf{29} (1996) 6795-6816
\bibitem{Draft}
P. Dorey, C. Dunning, A. Lishman and R. Tateo, ${\cal PT}$-symmetry breaking and exceptional points for a class of inhomogeneous complex potentials,  \textit{J. Phys. A} \textbf{42} (2009) 465302
\bibitem{Proof}
P. Dorey, C. Dunning and R. Tateo, Spectral equivalences, Bethe ansatz equations, and reality properties in ${\cal PT}$-symmetric quantum mechanics, \textit{J. Phys. A} \textbf{34} (2001) 5679-5704
\bibitem{Review}
P. Dorey, C. Dunning and R. Tateo, The ODE/IM correspondance, \textit{J. Phys. A} \textbf{40} (2007) R205
\bibitem{Anharmonic}
P. Dorey and R. Tateo, Anharmonic oscillators, the thermodynamic Bethe ansatz and nonlinear integral equations,  \textit{J. Phys. A} \textbf{32} (1999) L419-425
\bibitem{On}
P. Dorey and R. Tateo, On the relation between Stokes multipliers and the T-Q systems of conformal field theory,  \textit{Nucl. Phys. B} \textbf{563} (1999) 573-602
\bibitem{Tenszf}
E. Elizalde, Ten Physical Applications of Spectral Zeta Functions  (Springer 1995)
\bibitem{Alpha1}
V. A. Fateev and S. L. Lukyanov, Boundary RG flow associated with the AKNS soliton heirarchy, \textit{J. Phys. A} \textbf{39} (2006) 12889-12926
\bibitem{BLZKondo}
P. Fendley and H. Saleur, Exact perturbative solution of the Kondo problem,  \textit{Phys. Rev. Lett.} \textbf{75} (1995) 4492-4495
\bibitem{Grad}
I. S. Gradshteyn and I .M. Ryzhik, Table of Integrals, Series, and Products, 6th edn,  (New York: Academic 2000)
\bibitem{Zeta}
G. A. Mezincescu, Some properties of eigenvalues and eigenfunctions of the cubic oscillator with imaginary coupling constant, \textit{J. Phys. A} \textbf{33} (2000) 4911-4916
\bibitem{Newton}
R. G. Newton, The Complex j-Plane (New York: Benjamin 1964)
\bibitem{Shin}
K. C. Shin, The potential $(iz)^m$ generates real eigenvalues only,
under symmetric rapid decay conditions, \textit{J. Math. Phys} \textbf{46} (2005) 082110-082127
\bibitem{Sibuya}
Y. Sibuya, Global theory of a second-order linear ordinary differential equationwith polynomial coefficient, (Amsterdam: North-Holland 1975)
\bibitem{QES1}
A. V. Turbiner, Quasi-exactly-solvable problems and $\mathfrak{sl}(2)$ algebra, \textit{Commun. Math. Phys.} \textbf{118} (1988) 467-474
\bibitem{Vorosairy}
A. Voros, Airy function (exact WKB results for potentials of odd degree), \textit{\textit{J. Phys. A}} \textbf{32} (1999) 1301-1311
\bibitem{Vorosex}
A. Voros, Exercises in exact quantisation, \textit{J. Phys. A} \textbf{33} (2000) 7423-7450
\bibitem{Vorosspec0}
A. Voros, Spectral functions, special functions and the Selberg zeta function, \textit{Commun. Math. Phys.} \textbf{110} (1987) 439-465
\bibitem{Vorosspectral}
A. Voros, Spectral zeta functions, \textit{Adv. Stud. Pure Math.} \textbf{21} (1992) 327-358
\bibitem{Vorosquartic}
A. Voros, The return of the quartic oscillator. The complex WKB method, \textit{Ann. Inst. H. Poincar\'e} \textbf{39} (1983) 211-338.












\end{thebibliography}
\end{document}